\documentclass[11pt,a4paper]{article} 
\usepackage{jheppub}

\usepackage{amsmath, amssymb}
\usepackage{concmath, palatino}
\usepackage{mathrsfs}
\usepackage{array,arydshln}

\usepackage{graphicx,epsfig}
\usepackage{epic}
\usepackage{color}

\newcommand{\be}{\begin{equation}}
\newcommand{\ee}{\end{equation}}
\newcommand{\ba}{\begin{eqnarray}}
\newcommand{\ea}{\end{eqnarray}}

\newcommand{\mc}{\mathcal }

\newcommand{\N}{\mathcal{N}}

\def\XXint#1#2#3{{\setbox0=\hbox{$#1{#2#3}{\int}$}
     \vcenter{\hbox{$#2#3$}}\kern-.5\wd0}}

    \newcommand{\beq}{\begin{equation}}
    \newcommand{\eeq}{\end{equation}}
    \newcommand\beqa{\begin{eqnarray}}
    \newcommand\eeqa{\end{eqnarray}}

\newcommand{\brem}{Bremsstrahlung }

\title{On a discrete  symmetry of the Bremsstrahlung function in $\mc N=4$ SYM}

\author[a]{Matteo Beccaria } 
\author[b]{, Guido Macorini} 

\affiliation[a]{Dipartimento di Matematica e Fisica ìEnnio De Giorgi,\\
Universit\`a del Salento \& INFN, Via Arnesano, 73100 Lecce, 
Italy} 
                     
\affiliation[b]{Niels Bohr International Academy and Discovery Center,  \\
			Niels Bohr Institute, \\
		Blegdamsvej 17 DK-2100 Copenhagen, Denmark}

\emailAdd{matteo.beccaria@le.infn.it}
\emailAdd{macorini@nbi.ku.dk}

\abstract{
We consider the quark anti-quark potential on the three sphere in planar $\mc N = 4$ SYM
and the associated vacuum potential in the near BPS limit with $L$ units of $R$-charge. The associated \brem function 
$B_{L}$ has been recently computed analytically by means of the Thermodynamical Bethe Ansatz. We discuss it at
strong coupling by computing it at large but finite $L$. We provide strong support to a special symmetry of the \brem function under the formal discrete $\mathbb Z_{2}$ symmetry $L\to -1-L$. In this context, it is the counterpart of 
the reciprocity invariance discovered in the past in the spectrum of various gauge invariant composite operators.
The $\mathbb Z_{2}$ symmetry has remarkable consequences in the scaling limit where $L$ is taken to be large with
fixed ratio to the  't Hooft coupling. This limit organizes in inverse powers of the coupling and resembles
the semiclassical expansion of the dual string theory which is indeed known to capture the leading classical term.
We show that the various higher-order contributions to the \brem function obey several constraints and, in particular, 
the next-to-leading term, formally associated with the string one-loop correction, is completely determined by the classical contribution. The large $L$ limit at strong coupling is also discussed.
}

\allowdisplaybreaks

\begin{document} \maketitle

\bigskip

\section{Introduction and Results}

The application of integrability methods to realistic quantum field theories has a long history 
and started within the QCD context several decades ago \cite{Lipatov:1993yb,Faddeev:1994zg}.
Due to the paramount importance of AdS/CFT, it has been reconsidered somewhat later in planar $\mc N=4$ super Yang-Mills theory \cite{Minahan:2002ve}. From that moment, integrability has become a very common and useful tool to 
investigate strongly coupled gauge theories and their relation with string theory in the AdS/CFT perspective \cite{Beisert:2010jr}.

\medskip
On the gauge theory side of the correspondence, integrability allows a deep understanding of several observables
including the spectrum of gauge invariant composite operators, Wilson loops of various sorts, scattering amplitudes 
and correlation functions. In particular, the spectral problem is essentially solved non perturbatively due to the 
powerful machinery developed in \cite{Beisert:2006ez,Gromov:2009tv,Bombardelli:2009ns,Gromov:2009bc,Arutyunov:2009ur,Gromov:2009zb} among others.
Technically, the spectral problem is reformulated in terms of the Y-system which is an infinite set of functional equations supplemented with definite analytical properties \cite{Cavaglia:2010nm,Gromov:2011cx}.

\medskip
The very same methods that have been developed for the spectrum calculation of $\mc N=4$ SYM on $S^{3}$
can be applied to a different kind of observables related to the spectrum of the color flux between external quarks on $S^3$ \cite{Correa:2012hh,Drukker:2012de}. The vacuum energy of that flux is the same as the generalized cusp anomalous dimension $\Gamma_\text{cusp}$. This quantity has been introduced by Polyakov in \cite{Polyakov:1980ca}
and is the conformal dimension of a quark and anti-quark Wilson lines meeting at a cusp. Schematically, 
\be
\langle W\rangle=\left({\Lambda_\text{IR}\over\Lambda_\text{UV}}\right)^{\Gamma_\text{cusp}},
\ee
where $W$ is the cusped Wilson loop and $\Lambda_\text{UV, IR}$ are short and long distance cutoffs. The generalized
cusp anomalous dimension is a close relative of the conventional QCD cusp anomalous dimension 
\cite{Korchemsky:1985xj,Korchemsky:1988si}
that governs the scaling behavior of various gauge invariant quantities like logarithmic growth of the anomalous dimensions of high-spin
Wilson operators, Sudakov asymptotics of elastic form factors, the gluon Regge trajectory, infrared singularities
of on-shell scattering amplitudes and is one of the first observables computed at all orders in perturbation theory using integrability~\cite{Beisert:2006ez}.

\medskip
The cusp anomalous dimension is a function of two angles, $\phi$ and $\theta$ \cite{Drukker:1999zq}. 
The first angle, $\phi$, is the angle between the quark and antiquark lines meeting at the cusp. 
The second angle, $\theta$, characterizes the coupling to scalar fields in the locally supersymmetric Wilson lines. Indeed, the six real scalars of $\N=4$ SYM involve a coupling which is a unit vector $\vec n$ identifying a point of 
$S^5$. Thus, the quark-antiquark lines are associated to two different vectors, $\vec n$ and $\vec n_\theta$, with $\theta$ being the angle between them. Explicitly, we can write the cusped Wilson loop as $W_0 = W_{q}\,W_{\overline q}$,
with 
\be
W_{q} = {\rm P}\exp\!\int\limits_{-\infty}^0\! dt\left[i  A\cdot\dot{x}_q+\vec\Phi\cdot\vec n\,|\dot x_q|\right], \qquad
W_{\overline q} =  {\rm P}\exp\!\int\limits_0^\infty\!dt\left[i A\cdot\dot x_{\bar q}+\vec\Phi\cdot\vec n_{\theta}\,|\dot x_{\bar q}|\right],
\ee
where $\vec \Phi$ denotes a vector consisting of the six scalars of $\N=4$ SYM, while $x_q(t)$ and $x_{\bar q}(t)$ are the quark and antiquark trajectories, i.e. straight lines through the origin, making up the angle $\phi$ at the cusp.

\medskip
The existence of the two parameters $\phi, \theta$ allows  to consider various special limits. For instance, when $\phi^2-\theta^2\to0$ and also $\phi^2=\theta^2$ the cusped Wilson loop is BPS and the energy vanishes as shown in \cite{Zarembo:2002an}. If instead, $\theta=0$ and $\phi\to 0$, the cusp anomaly
behaves as ($g=\frac{\sqrt\lambda}{4\pi}$ in terms of the 't Hooft coupling $\lambda$)
\be
\Gamma_\text{cusp}(g; \phi,0)  = -\phi^2 B_0(g) + {\cal O}\left(\phi^4\right)\;.
\ee
where the function $B_0$ is known as the {\it Bremsstrahlung function} \cite{Correa:2012at,Fiol:2012sg}
and controls the power radiation of an accelerating quark. The Bremsstrahlung function has been computed exactly in \cite{Correa:2012at,Fiol:2012sg} using results from 
localization \cite{Bassetto:2008yf,Bassetto:2009ms,Bassetto:2009rt,Erickson:2000af,Drukker:2000rr,Drukker:2006ga,Drukker:2006zk,Drukker:2007dw,Drukker:2007yx,Giombi:2009ms,Giombi:2009ds,Pestun:2007rz,Pestun:2009nn}. In the planar limit, it reads
\be
\label{eq:local}
B_0(g) =g^2\left(1-\frac{I_3(4\pi g)  }{I_1(4\pi g) }\right)  
\ee
where $I_n$ are modified Bessel functions of the first kind.

\medskip
A non-perturbative description of  $\Gamma_\text{cusp}$ has been obtained in \cite{Drukker:2012de,Correa:2012hh}
where an infinite system of thermodynamical Bethe Ansatz (TBA) integral equations has been derived to 
compute $\Gamma_\text{cusp}$  at any  coupling  $g$ and  angles $\phi, \theta$. 
The TBA formulation actually considers a generalization of the described set-up where 
a local operator with R-charge $L$ is inserted at the cusp 
\beq
W_L=W_{q}\cdot Z^{L}\cdot W_{\overline q},
\eeq
where $Z=\Phi_1+i\Phi_2$, with $\Phi_1$ and $\Phi_2$ being two scalars independent from $\vec\Phi\cdot\vec n$ 
and $\vec\Phi\cdot\vec n_\theta$. The associated anomalous dimension $\Gamma_L(g; \phi,\theta)$ 
 is computed as the vacuum energy by the TBA equations exactly at any value of the parameter $L$ which plays the role of the system size in the TBA language. 
For $L=0$, one recovers the usual quark-antiquark potential, see for instance \cite{Drukker:2011za}. From $\Gamma_L(g; \phi,\theta)$, one can go in the above small angle limits ($\phi\to 0, \theta=0$) and define a generalized \brem function 
$B_{L}(g)$.

\medskip
In the remarkable paper \cite{Gromov:2012eu}, the function $B_{L}(g)$ has been computed analytically for all $g$ and $L$ by exploiting the relevant simplifications that occur in the TBA equations in the small angle limit. 
Later, in \cite{Gromov:2013qga}, the method has been extended to cover the case of a non zero $\theta$ angle.
This is non trivial and is based on the reduction of the TBA problem to a finite set of equations, known as 
FiNLIE \cite{Gromov:2011cx}, as well as the very recent further reduction called ${\bf P}\mu$-system \cite{Gromov:2013pga}.

\medskip
In this short paper, we build on the results of \cite{Gromov:2012eu,Gromov:2013qga} and analyze a special property
of the \brem function at strong coupling that has  interesting consequences. 
Let us first consider the $\theta=0$ case for simplicity. An intriguing remark in \cite{Gromov:2012eu} is 
the fact that at strong coupling the \brem function appear to admit the following structure
\be
\label{eq:strong}
B_{L}(g) = \sum_{p=0}^{\infty}\frac{g^{1-p}}{\pi^{p+1}}P_{p}(L), 
\ee
where $P_{p}(L)$ are  degree $p$ polynomials $P_{p}(L)=\sum_{n=0}^{p}c_{p,n}L^{n}$. 
This means that it is sensible to consider the scaling limit $g\to \infty$ with fixed ratio
$\mc L = L/g$. The \brem function admits then the following expansion 
\be
\label{eq:semiclassical}
\mc B(g; \mc L) \equiv  \frac{1}{g}\,B_{\mc L\,g}(g) = \sum_{q=0}^{\infty}\frac{1}{g^{p}}\,b_{p}(\mc L),
\ee
where the functions $b_{p}(\mc L)$ admit a regular expansion around $\mc L=0$,
\be
b_{p}(\mc L) = \sum_{n=0}^{\infty}b_{p,n}\,\mc L^{n},\qquad
b_{p,n} = \frac{1}{\pi^{n+p+1}}c_{n+p,n}.
\ee
This kind of expansions reminds the similar semiclassical expansion in the dual string theory
in terms of the semiclassical charges that are scaled by $g$ (see for instance
the discussion in \cite{Beccaria:2012xm}).
The classical term $b_{0}(\mc L)$ involves (all) the leading coefficients $c_{n,n}$ and can be computed exactly by solving the equations of the dual classical string theory
as in \cite{Correa:2012hh} with full agreement. Similarly, it is very tempting to 
associate the functions $b_{p}(\mc L)$ to higher loop corrections in string theory. This approach is definitely non trivial
since it compares two different orders of limits (large $L$ and large $g$).  It may encounter obstructions and be only partially valid depending on the degree of protection of the involved coefficients \cite{Beccaria:2012kp}. With these remarks being understood, 
we shall be calling $b_{p}(\mc L)$ a $p$-th loop semiclassical contribution although
its precise relation with the would be $b^{\rm string}_{p}(\mc L)$ function requires a detailed comparison.

\medskip
The simplest hint that some non-renormalization properties are at work  is provided by 
the one-loop term $b_{1}(\mc L)$ that involves all subleading coefficients $c_{n,n-1}$.
Its leading term at small $\mc L$ is $b_{1,0}$ and this coefficient turns out to be  in agreement with the world-sheet
explicit one-loop string calculation of \cite{Drukker:2011za}. In principle, the complete determination of $b_{1}(\mc L)$ could be attempted by extending the methods of that paper. Here, we stay in the gauge theory and show that $b_{1}(\mc L)$
is indeed fully determined by the classical term. Actually, there is a full set of constraints that allow to express all 
odd functions $b_{2n+1}(\mc L)$ in terms of the even ones $b_{2n}(\mc L)$. The constraints read
\ba
\label{eq:relation}
b_{1}(\mc L) &=& \frac{1}{2}\frac{d}{d\mc L}b_{0}(\mc L), \nonumber \\
b_{3}(\mc L) &=& \frac{1}{2}\frac{d}{d\mc L}b_{2}(\mc L)-\frac{1}{24}\frac{d^{3}}{d\mc L^{3}}b_{0}(\mc L), \\
b_{5}(\mc L) &=& \frac{1}{2}\frac{d}{d\mc L}b_{4}(\mc L)-\frac{1}{24}\frac{d^{3}}{d\mc L^{3}}b_{2}(\mc L)+
\frac{1}{240}\frac{d^{5}}{d\mc L^{5}}b_{0}(\mc L), \nonumber
\ea
and so on. In particular, the one-loop function is trivially computable from the classical $b_{0}(\mc L)$.
This surprising result is a consequence of a strong coupling $\mathbb Z_{2}$ symmetry of $B_{L}(g)$ that can be written~\footnote{One could look for a similar relation in the exact formula for the so-called {\em slope} describing
the small spin limit of the minimal scaling dimension of twist $L$ Wilson operators in the $\mathfrak{sl}(2)$ sector of 
the planar $\mathcal N = 4$ SYM theory \cite{Basso:2011rs,Basso:2012ex,Gromov:2012eg}.
However, that expression has a $L$ dependence which is much simpler
than that of $B_{L}$ and no special symmetry in $L$ can be found.}
\be
\label{eq:reciprocity}
\phantom{\frac{1}{1}}
B_{L}(g) = -B_{-1-L}(-g)\;.
\ee
In principle, the parameter $L$ is a non negative integer, but it is clear that the structure (\ref{eq:strong}) allows for a 
non ambiguous continuation to negative integers.
This relation loosely reminds an analogous invariance known as reciprocity invariance. It appears 
in the spectrum calculation of  various anomalous dimensions of gauge 
invariant composite operators of $\mc N=4$ SYM \cite{Basso:2006nk} (see also \cite{Beccaria:2010tb} and references therein). At weak coupling, the leading term of $B_{L}(g)$ is the leading L\"uscher correction computed in \cite{Correa:2012at}. After some manipulations, it can be written as $2g^{2L+2}(2^{2L}-2)\zeta_{2L}$. This form allows for analytic continuation in $L$, but no special symmetry is apparent.

\medskip
The plan of the paper is the following. In Sec.~(\ref{sec:definitions}), we give some preliminary definitions.
In Sec.~(\ref{sec:efficient}), we show how to efficiently compute $B_{L}$ at strong coupling. 
In Sec.~(\ref{sec:results}), we present our main results, including a discussion of the $\mathbb Z_{2}$ symmetry. Finally, 
in Sec.~(\ref{sec:application}), we discuss the relations (\ref{eq:relation}).

\section{Preliminary definitions}
\label{sec:definitions}

According to the results in \cite{Gromov:2012eu}, the \brem function $B_{L}(g)$ is 
\be
B_L=g^2(-R_{L+1}+2R_L-R_{L-1})\;,
\ee
where $R_{L}$ is the following ratio of determinants
\be
R_L\equiv
\left|
\begin{array}{ccccc}
I_{1}&I_{3}&\dots&I_{2L-1}&I_{2L+1}\\ \hdashline
I_{-3}&I_{-1}&\dots&I_{2L-5}&I_{2L-3}\\
\vdots&\vdots& \ddots & \vdots& \vdots\\
I_{1-2L}&I_{3-2L}&\dots&I_{-1}&I_{1}\\
I_{-1-2L}&I_{1-2L}&\dots&I_{-3}&I_{-1}
\end{array}
\right|/
\left|
\begin{array}{ccccc}
I_{-1}&I_{1}&\dots&I_{2L-3}&I_{2L-1}\\
I_{-3}&I_{-1}&\dots&I_{2L-5}&I_{2L-3}\\
\vdots&\vdots& \ddots & \vdots& \vdots\\
I_{1-2L}&I_{3-2L}&\dots&I_{-1}&I_{1}\\
I_{-1-2L}&I_{1-2L}&\dots&I_{-3}&I_{-1}
\end{array}
\right|\;,
\ee
with $I_{n}\equiv I_n(4\pi g)$. In particular, at $L=0$, we immediately recover the result from localization
(\ref{eq:local}).
The large $g$ expansion of $B_{L}$ takes the form (\ref{eq:strong}) with the 
following polynomials reported in \cite{Gromov:2012eu}
\ba
P_{0}(L) &=& 1, \\
P_{1}(L) &=& -\frac{1}{8}(6L+3), \\
P_{2}(L) &=& \frac{3}{128}(6L^{2}+6L+1), \\
P_{3}(L) &=& \frac{1}{512}(10L^{3}+15L^{2}+11L+3), \\
P_{4}(L) &=& -\frac{9}{32768}(10L^{4}+20L^{3}-22L^{2}-32L-7).
\ea
Setting $\mc L = L/g$, the large $g$ expansion of $B_{g\,\mc L}(g)$ takes the form (\ref{eq:semiclassical}).
The classical function $b_{0}(\mc L)$ can be found by eliminating $q$ between the two equations
\ba
\label{eq:b0-param}
b_{0} &=& \frac{1-q}{2\,\mathbb E(q)}, \nonumber \\
\mc L &=& 4\,(\mathbb K(q)-\mathbb E(q)).
\ea
This leads to~\footnote{Alternatively, by the methods of \cite{Pawellek:2011xd}
one can show that  the function $b_{0}(\mc L)$ obeys the following differential equation
\ba
&& -\mc L b_{0}^{3}(\mc L b_{0}-4){b_{0}''}^{2}+2b_{0}b_{0}''(-3\mc L b_{0}b_{0}'+4b_{0}'+3b_{0}^{2})
(-\mc L b_{0}b_{0}'+2b_{0}'+2b_{0}^{2})\nonumber \\
&&\qquad\qquad -{b_{0}'}^{2}(-3\mc L b_{0} b_{0}'+4b_{0}'+3b_{0}^{2})^{2}=0
\ea
with boundary conditions $b_{0}(0) = \frac{1}{\pi}$, $b_{0}'(0) = -\frac{3}{4\pi^{2}}$.}
\be
b_{0}(\mc L) = \frac{1}{\pi }-\frac{3 \mathcal{L}}{4 \pi ^2}+\frac{9 \mathcal{L}^2}{64 \pi ^3}+\frac{5 \mathcal{L}^3}{256 \pi ^4}-\frac{45 \mathcal{L}^4}{16384 \pi ^5}-\frac{63 \mathcal{L}^5}{32768 \pi
   ^6}-\frac{245 \mathcal{L}^6}{1048576 \pi ^7}+\frac{459 \mathcal{L}^7}{4194304 \pi ^8}+\cdots .
\ee

The higher loop contributions $b_{n}(\mc L)$ with $n>0$ can be computed by re-expanding the expression of $B_{L}$
and in particular, one finds in \cite{Gromov:2012eu} the one-loop term
\be
b_{1}(\mc L) = -\frac{3}{8\pi^{2}}+\frac{9}{64\pi^{3}}\mc L+\frac{15}{512\pi^{4}}\mc L^{2}-\frac{45}{8192\pi^{5}}\mc L^{3}+\mc O(\mc L^{4}),
\ee
whose first term agrees with the string calculation of \cite{Drukker:2011za}.

\section{Efficient calculation of $B_{L}$ at strong coupling}
\label{sec:efficient}

An efficient algorithm to compute $B_{L}$ at strong coupling is based on the following three elementary but useful steps.

\begin{enumerate}
\item[a)] {\bf Reduction to a linear problem}. 
Let $A$ be an invertible matrix $A = (a_{ij})$. Let $\widehat A$ be a matrix with the same dimension 
such that $\widehat A_{ij}=a_{ij}$ for $i\neq 1$ and $\widehat A_{1j} = \widehat a_{j}$. So, we changed the first row.
Then,
\be
\frac{\det \widehat A}{\det A} = v_{1},
\ee
where $v_{i}$ is the solution to 
\be
A^{T}\,v = \widehat a.
\ee

\item[b)] {\bf Identification of the non exponentially suppressed terms at large $g$}.
At large $g$, we can replace up to exponentially depressed terms 
\be
\label{eq:rep}
I_{n}(4\pi g) \longrightarrow \frac{1}{2\,\pi\,\sqrt{2g}}\,{}_{2}F_{0}
\left(n+\frac{1}{2}, \frac{1}{2}-n; \frac{1}{8\,\pi\,g}\right).
\ee
The meaning of the r.h.s for positive integer $n$ is just that of a bookkeeping for the expansion around $g=\infty$.

\item[c)] {\bf Iterative expansion of the linear problem in inverse powers of $g$}.
The large $g$ expansion can be worked out systematically by the standard methods used in quantum mechanical degenerate perturbation theory.
Let us denote for simplicity $A^{T}=S$ (standing for singular) ,  $x = 4\,\pi\,g$, and $\varepsilon = \frac{1}{\sqrt x}$. Also, let us make the replacement (\ref{eq:rep}) everywhere. The linear problem
\be
S(\varepsilon)\,v(\varepsilon) = \widehat a(\varepsilon),
\ee
can be formally expanded in powers of $\varepsilon$. The expansion takes the form 
\ba
&& \left[\varepsilon\,S^{(1)}+\varepsilon^{3}\,S^{(3)}+\varepsilon^{5}\,S^{(5)}+\cdots\right]\,\left[
v^{(0)}+\varepsilon^{2}\,v^{(2)}+\varepsilon^{4}\,v^{(4)}+\cdots\right]= \nonumber \\
&& \qquad\qquad=\varepsilon \, \widehat a^{(1)}+\varepsilon^{3}\,\widehat a^{(3)}+\varepsilon^{5}\,\widehat a^{(5)}+\cdots\, .
\ea
This means 
\ba
S^{(1)}\,v^{(0)} &=& \widehat a^{(1)}, \\
S^{(1)}\,v^{(2)} &=& \widehat a^{(3)}-S^{(3)}\,v^{(0)}, \\
S^{(1)}\,v^{(4)} &=& \widehat a^{(5)}-S^{(5)}\,v^{(0)}-S^{(3)}\,v^{(2)}, \\
&\cdots&\nonumber
\ea
The matrix $S^{(1)}$ is a $(L+1)\times (L+1)$ matrix with constant coefficients $S^{(1)}_{ij} = \frac{1}{\sqrt{2\pi}}$.
Its eigenvectors are $(1, \dots, 1)^{T}$ with eigenvalue $\frac{L+1}{\sqrt{2\pi}}$ and a $L$ dimensional null space
$\mathbb V_{0}$ with eigenvectors
\be
\nu_{i} = (1, 0, \dots, 0, -1, 0, \dots 0),\quad i=2, \dots, L+1,
\ee
where $-1$ is in the $i$-th position. Let us denote by $P_{0}$ the projector onto $\mathbb V_{0}$. The first equation can be solved iff $P_{0}\widehat a^{(1)}=0$ and this turns out to be true. Then, 
\be
v^{(0)} = \frac{\sqrt{2\pi}}{L+1}\,\widehat a^{(1)}+\nu^{(0)},
\ee
where $\nu^{(0)}$ is a generic vector in $\mathbb V_{0}$. It is partially fixed by the necessary (linear) condition 
\be
P_{0}(\widehat a^{(3)}-S^{(3)}\,v^{(0)})=0.
\ee
Imposing this condition, one can solve $v^{(2)}$ as
\be
v^{(2)} = \frac{\sqrt{2\pi}}{L+1}\,(\widehat a^{(3)}-S^{(3)}\,v^{(0)})+\nu^{(2)},
\ee
where, again, $\nu^{(2)}\in \mathbb V_{0}$. Going on in this way, one efficiently determines the expansion 
vectors $v^{(n)}$ and extracts in the end its first component $v(\varepsilon)_{1}$.

\end{enumerate}

\section{Results and $\mathbb Z_{2}$ invariance}
\label{sec:results}

By means of the methods that we have presented it is easy to compute the polynomials $P_{p}(L)$ at strong coupling for 
increasing $L$. In this way, we have confirmed the general structure in (\ref{eq:strong}) by computing explicitly
the polynomials for several values of $L$. The first cases, extending the results of \cite{Gromov:2012eu},
are:
\ba
P_{5}(L) &=& -\frac{9}{2^{16}}(14L^{5}+35L^{4}+20L^{3}-5L^{2}-16L-6), \\
P_{6}(L) &=& -\frac{1}{2^{22}}(980L^{6}+2940L^{5}+11570L^{4}+18240L^{3}+1436L^{2}-7194L-1899)\\
P_{7}(L) &=&\frac{27}{2^{23}}(34L^{7}+119L^{6}-189L^{5}-770L^{4}-679L^{3}-189L^{2}+186L+96)\\
P_{8}(L) &=& \frac{3}{2^{31}} (35010 L^8+140040 L^7+212660 L^6+147840 L^5-1024690 L^4-2132400 L^3-447484 L^2\nonumber \\
&& +620016 L+181161), \\
P_{9}(L) &=& \frac{(2 L+1)}{2^{32}} (4565 L^8+18260 L^7+479258 L^6+1373864 L^5-526555 L^4\nonumber \\
&& -3321580 L^3-2279988 L^2-393264 L+1037664),\\
P_{10}(L) &=& -\frac{9}{2^{38}} (129052 L^{10}+645260 L^9-991470 L^8-7837440 L^7-17720024 L^6\nonumber \\
&& -23018940 L^5+34384350 L^4+96699400 L^3+24449206 L^2-26237166 L-8027901), \\
P_{11}(L) &=& -\frac{3 (2 L+1)}{2^{36}} (15132 L^{10}+75660 L^9+267035 L^8+614180 L^7-3483880 L^6\nonumber \\
&& -12283498 L^5+215495 L^4+21468710 L^3+16676658 L^2+3815388 L-7340328), \\
P_{12}(L) &=& \frac{3}{2^{46}} (1836380 L^{12}+11018280 L^{11}-344159948 L^{10}-1821800640 L^9-816960390 L^8\nonumber \\
&& +7784163360 L^7+24854488456 L^6+39546129840 L^5-34325714980 L^4\nonumber \\
&& -121775920620 L^3-33316523250
   L^2+32309376048 L+10137685113), \\
P_{13}(L) &=& \frac{9 (2 L+1)}{2^{45}} (305725 L^{12}+1834350 L^{11}-2587441 L^{10}-29752080 L^9-161161525 L^8\nonumber \\
&& -445955770 L^7+1028773081 L^6+4502027852 L^5+561988080 L^4-6830092240 L^3\nonumber \\
&& -5716882080 L^2-1487626272
   L+2495860992), \\
P_{14}(L) &=& \frac{9}{ 2^{53}}(33816600 L^{14}+236716200 L^{13}+1978360748 L^{12}+8792853888 L^{11}\nonumber \\
&& -30911443640 L^{10}-229516642740 L^9-224256476166 L^8+567123796944 L^7\nonumber \\
&& +2535219437800 L^6+4570188856800
   L^5-2901336130672 L^4-12255744128712 L^3\nonumber \\
   && -3504907996632 L^2+3214738115658 L+1025393923683), 
\ea
and so on. From these expressions, we derive the various $b_{p}(\mc L)$ functions that read
\ba
\label{eq:b0}
b_{0}(\mc L) &=& \frac{1}{\pi }-\frac{3 \mathcal{L}}{4 \pi ^2}+\frac{9 \mathcal{L}^2}{64 \pi ^3}+\frac{5 \mathcal{L}^3}{256 \pi ^4}-\frac{45 \mathcal{L}^4}{16384 \pi ^5}-\frac{63 \mathcal{L}^5}{32768 \pi
   ^6}\nonumber \\
   && -\frac{245 \mathcal{L}^6}{1048576 \pi ^7}+\frac{459 \mathcal{L}^7}{4194304 \pi ^8}+\frac{52515 \mathcal{L}^8}{1073741824 \pi ^9}+O\left(\mathcal{L}^9\right), \\
   %%%
   \label{eq:b1}
   b_{1}(\mc L) &=& -\frac{3}{8 \pi ^2}+\frac{9 \mathcal{L}}{64 \pi ^3}+\frac{15 \mathcal{L}^2}{512 \pi ^4}-\frac{45 \mathcal{L}^3}{8192 \pi ^5}-\frac{315 \mathcal{L}^4}{65536 \pi ^6}-\frac{735 \mathcal{L}^5}{1048576
   \pi ^7}+\frac{3213 \mathcal{L}^6}{8388608 \pi ^8}\nonumber \\
   &&+\frac{52515 \mathcal{L}^7}{268435456 \pi ^9}+\frac{41085 \mathcal{L}^8}{4294967296 \pi ^{10}}+O\left(\mathcal{L}^9\right), \\
   %%%
   \label{eq:b2}
   b_{2}(\mc L) &=& \frac{3}{128 \pi ^3}+\frac{11 \mathcal{L}}{512 \pi ^4}+\frac{99 \mathcal{L}^2}{16384 \pi ^5}-\frac{45 \mathcal{L}^3}{16384 \pi ^6}-\frac{5785 \mathcal{L}^4}{2097152 \pi ^7}-\frac{5103
   \mathcal{L}^5}{8388608 \pi ^8}+\frac{159495 \mathcal{L}^6}{536870912 \pi ^9}\nonumber \\
   &&+\frac{122097 \mathcal{L}^7}{536870912 \pi ^{10}}+\frac{4461615 \mathcal{L}^8}{137438953472 \pi
   ^{11}}+O\left(\mathcal{L}^9\right), \\
   %%%
   \label{eq:b3}
   b_{3}(\mc L) &=&\frac{3}{512 \pi ^4}+\frac{9 \mathcal{L}}{1024 \pi ^5}+\frac{45 \mathcal{L}^2}{65536 \pi ^6}-\frac{285 \mathcal{L}^3}{65536 \pi ^7}-\frac{10395 \mathcal{L}^4}{4194304 \pi ^8}+\frac{3465
   \mathcal{L}^5}{16777216 \pi ^9}+\frac{1613493 \mathcal{L}^6}{2147483648 \pi ^{10}}\nonumber \\
   &&+\frac{275535 \mathcal{L}^7}{1073741824 \pi ^{11}}-\frac{4486185 \mathcal{L}^8}{68719476736 \pi
   ^{12}}+O\left(\mathcal{L}^9\right), \\
   %%%
   \label{eq:b4}
   b_{4}(\mc L) &=&\frac{63}{32768 \pi ^5}+\frac{9 \mathcal{L}}{4096 \pi ^6}-\frac{359 \mathcal{L}^2}{1048576 \pi ^7}-\frac{18333 \mathcal{L}^3}{8388608 \pi ^8}-\frac{1537035 \mathcal{L}^4}{1073741824 \pi
   ^9}+\frac{160377 \mathcal{L}^5}{2147483648 \pi ^{10}}\nonumber \\
   && +\frac{19935027 \mathcal{L}^6}{34359738368 \pi ^{11}}
   +\frac{4765185 \mathcal{L}^7}{17179869184 \pi ^{12}}-\frac{1225440585
   \mathcal{L}^8}{35184372088832 \pi ^{13}}+O\left(\mathcal{L}^9\right), \\
   %%%
   \label{eq:b5}
   b_{5}(\mc L) &=&\frac{27}{32768 \pi ^6}+\frac{3597 \mathcal{L}}{2097152 \pi ^7}-\frac{5103 \mathcal{L}^2}{8388608 \pi ^8}-\frac{399825 \mathcal{L}^3}{134217728 \pi ^9}-\frac{7169715 \mathcal{L}^4}{4294967296 \pi
   ^{10}}+\frac{51792615 \mathcal{L}^5}{68719476736 \pi ^{11}}\nonumber \\
   && +\frac{21038157 \mathcal{L}^6}{17179869184 \pi ^{12}}+\frac{729765315 \mathcal{L}^7}{2199023255552 \pi ^{13}}-\frac{9477657585
   \mathcal{L}^8}{35184372088832 \pi ^{14}}+O\left(\mathcal{L}^9\right).
\ea
From these long expansions, we immediately observe an intriguing relation between the one-loop term and the 
classical one, namely
\be
\label{eq:remark}
b_{1}(\mc L) = \frac{1}{2}\,\frac{d}{d\mc L}b_{0}(\mc L).
\ee
This is definitely non trivial. For instance,  it implies that the complicated string calculation in \cite{Drukker:2011za}
determining $b_{1}(0)$ could be simply replaced by the classical term $b_{0}'(0)$. The question is then: Is there any 
reason behind (\ref{eq:remark}) or is it an accident ? To answer this question, we analyzed the polynomials 
$P_{p}(L)$  and found in all cases the following remarkable property
\be
\label{eq:Z2p}
P_{p}(L) = (-1)^{p}\,P_{p}(-1-L)\;,
\ee
which is a non trivial discrete symmetry of the \brem function. Of course, this is the same as the compact relation
(\ref{eq:reciprocity}) to be meant in the large $g$ expansion. The relation between (\ref{eq:Z2p}) and (\ref{eq:remark})
is elucidated in the next section.

\section{Simple application: Constraints on the $b_{p}(\mc L)$ functions}
\label{sec:application}

The $\mathbb Z_{2}$ symmetry (\ref{eq:Z2p}) implies the following structure of the polynomials $P_{p}(L)$. Let us define the shifted polynomials
$\widetilde P_{p}(U) = P_{p}(U-\frac{1}{2})$. Then
\be
\widetilde P_{p}(-U) = P_{p}(-U-\frac{1}{2}) = P_{p}(-1-(U-\frac{1}{2})) = 
(-1)^{p} P_{p}(U-\frac{1}{2}) = (-1)^{p}\,\widetilde P_{p}(U).
\ee
In other words, $\widetilde P_{p}(U)$ contains only even or odd powers of $U$. In terms of $L$ this means that
half of the coefficients of $P_{p}(L)$ are determined by the other ones because $P_{p}(L)$  can be written as a polynomial in $L+1/2$ with only even or odd coefficients
\be
P_{p}(L) = c_{p}\bigg(L+\frac{1}{2}\bigg)^{p}+c_{p-2}\bigg(L+\frac{1}{2}\bigg)^{p-2}
+c_{p-4}\bigg(L+\frac{1}{2}\bigg)^{p-4}+\cdots.
\ee
Expanding the various monomials, we immediately deduce the  relations (\ref{eq:relation}) that can be checked on the 
results (\ref{eq:b0}-\ref{eq:b5}).

\subsection{Large $\mc L$ limit}

\begin{figure}
\begin{center}
\includegraphics[width=0.49\textwidth]{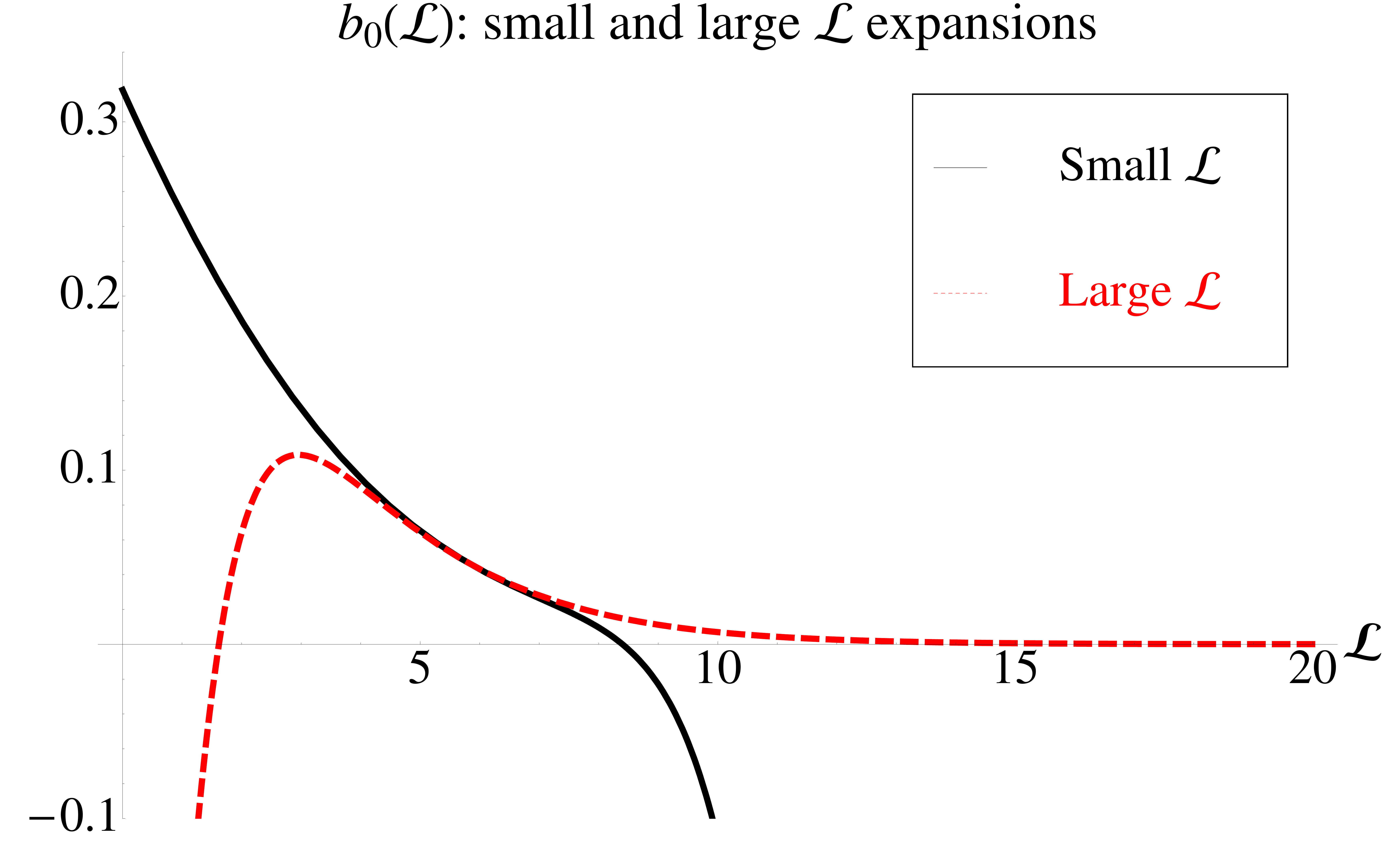}
\includegraphics[width=0.49\textwidth]{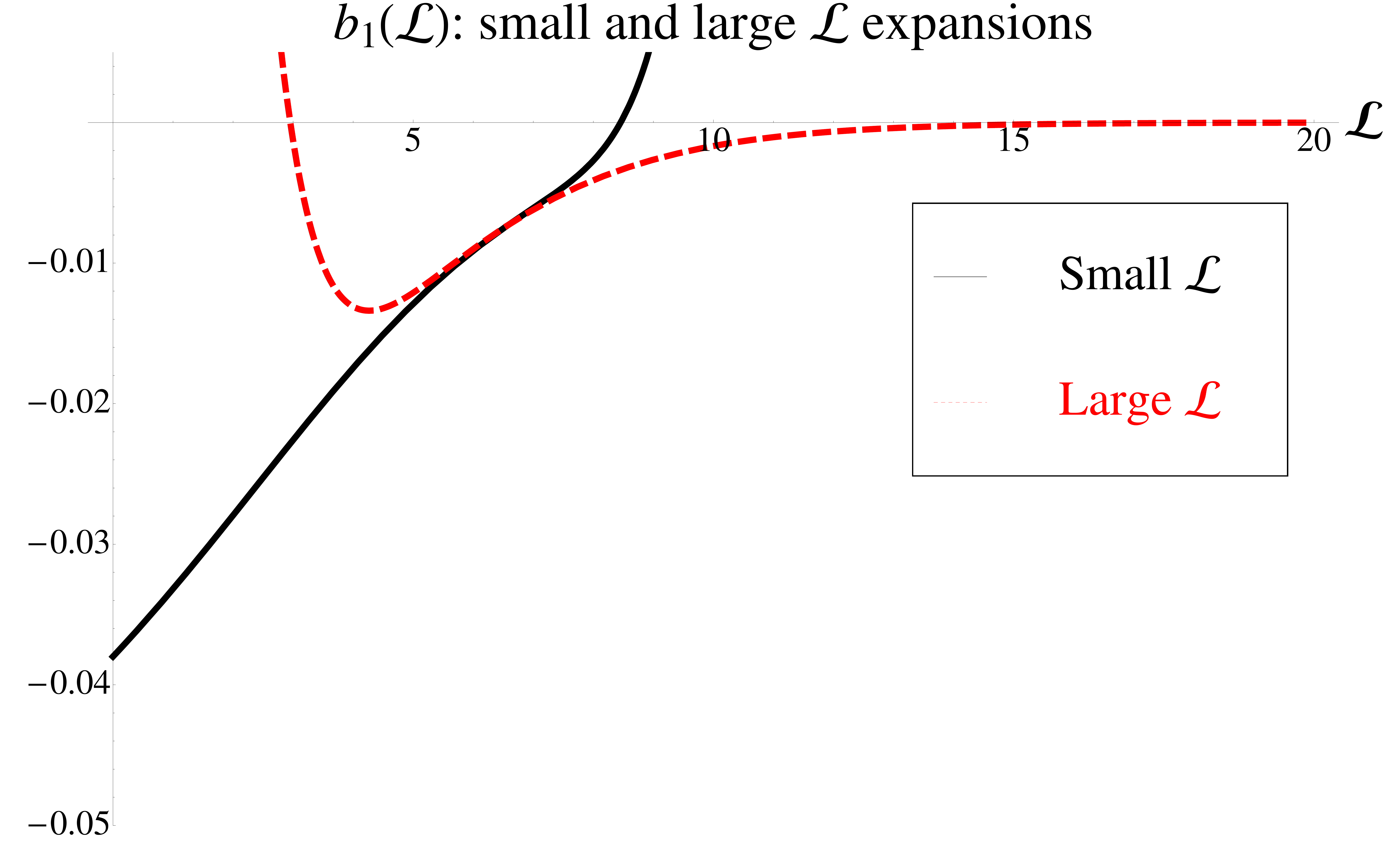}
\end{center}
\caption{Large and small $\mathcal{L}$ expansions for the
$b_0(\mathcal{L})$ and $b_1(\mathcal{L})$ functions. The weak coupling curve is the expansion up to $\mc O(\mc L^{12})$.}
\label{fig:largesmallL}
\end{figure}

\begin{figure}
\begin{center}
\includegraphics[width=0.49\textwidth]{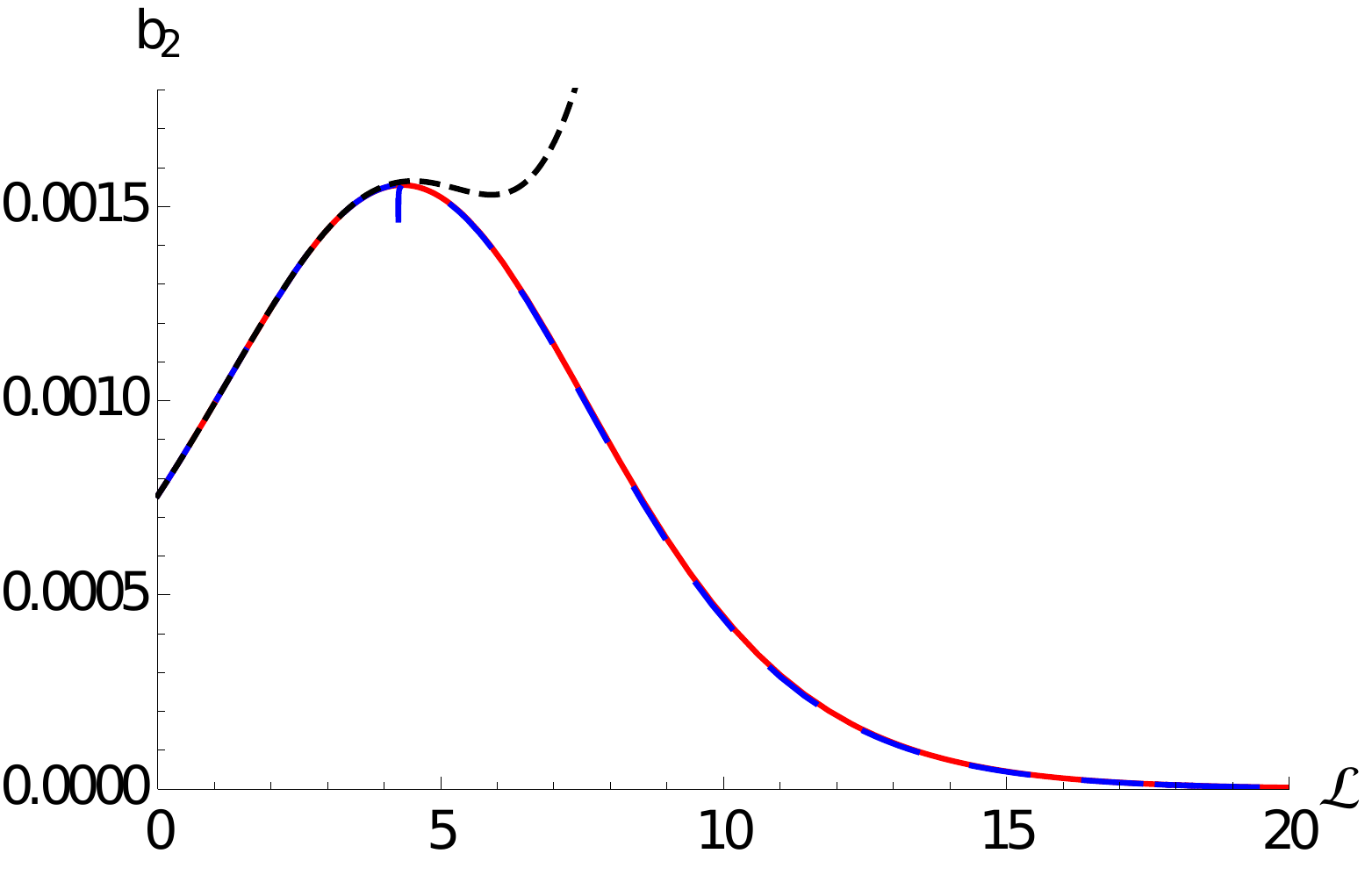}
\includegraphics[width=0.49\textwidth]{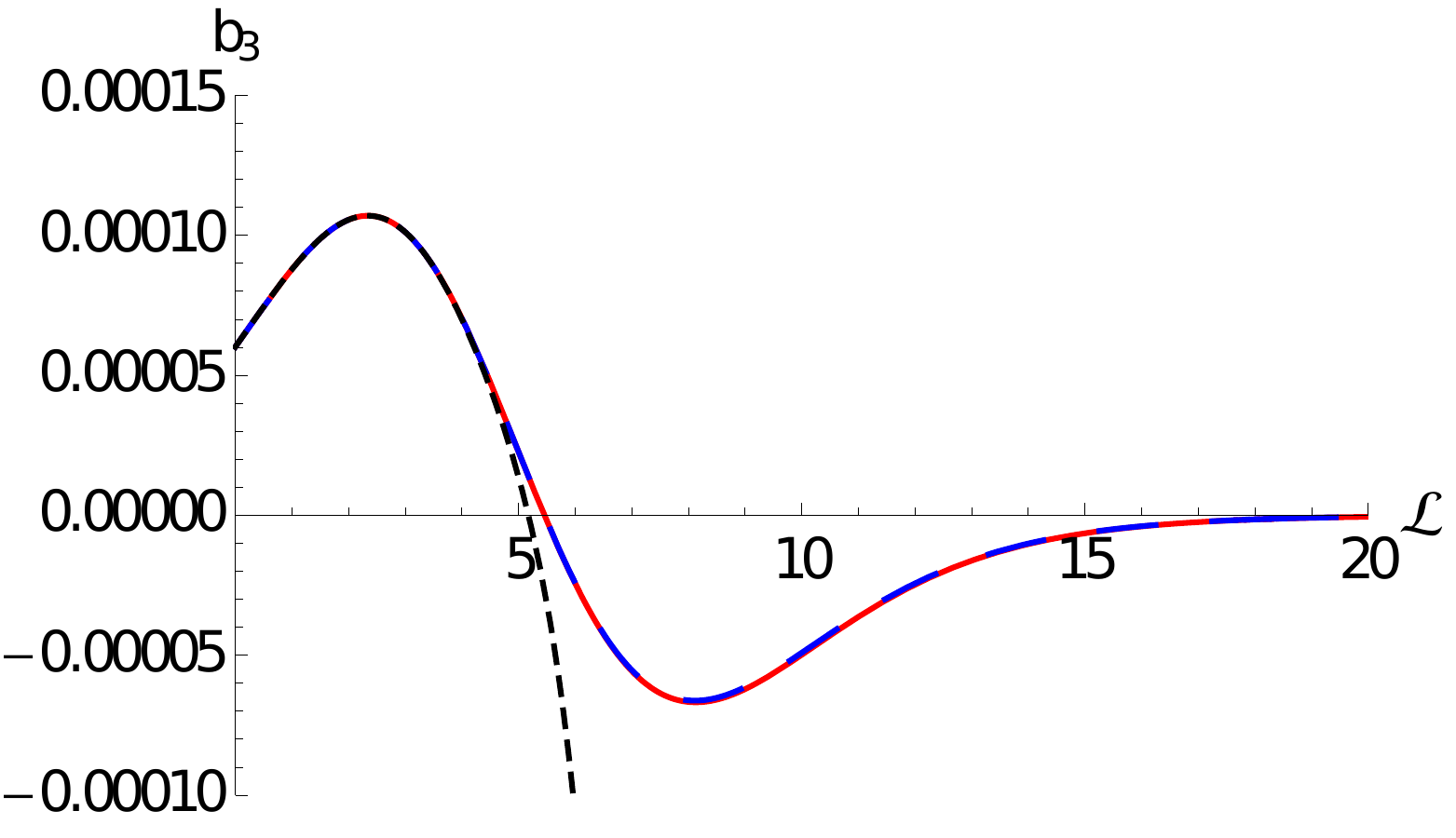}
\end{center}
\caption{Pad\'e resummation of the functions $b_{2,3}(\mathcal{L})$ as described in the main text.
The dashed black line is the weak coupling expansion. The solid red line is obtained with $[N, M]=[5,4]$. The dashed blue line with $[N,M]=[6,5]$ and is numerically coincident with the previous one.}
\label{fig:pade}
\end{figure}

The parametric form of  $b_{0}(\mc L)$ given in (\ref{eq:b0-param}) allows to explore the large $\mc L$ limit by 
going to $q\to 1$. One finds the following expansion 
\be
b_{0}(\mc L) = \frac{\xi }{2}+\xi ^2 \left(-\frac{\mathcal{L}}{8}-\frac{3}{8}\right)+\frac{1}{256} \xi ^3 \left(9 \mathcal{L}^2+39 \mathcal{L}+56\right)+\frac{\xi ^4
   \left(-16 \mathcal{L}^3-84 \mathcal{L}^2-198 \mathcal{L}-183\right)}{1536}+\cdots, 
   \ee
in terms of the exponentially small parameter
\be
\xi = 16\,e^{-\frac{1}{2}\mc L-2}.
\ee
The first term of the expansion reproduces the result of \cite{Correa:2012hh} for the leading L\"uscher correction
at strong coupling.
The same expansion can be computed for the one-loop function $b_{1}(\mc L)$ since its series expansion 
at $\mc L=0$ can be resummed using (\ref{eq:reciprocity}). The result is 
\be
b_{1} = \frac{(q-1)\left[(q-1)\,\mathbb K(q)+(q+1)\,\mathbb E(q)\right]}{16\,q\,\mathbb E(q)^{3}},
\ee
where, again $q$ is related to $\mc L$ by means of the second of (\ref{eq:b0-param}). Expanding around $q=1$
or simply using $b_{1}(\mc L) = \frac{1}{2} b_{0}'(\mc L)$ we find
\be
b_{1}(\mc L) = -\frac{\xi }{8}+\frac{1}{16} \xi ^2 (\mathcal{L}+2)-\frac{9 \xi ^3 \left(3 \mathcal{L}^2+9 \mathcal{L}+10\right)}{1024}+\frac{1}{768} \xi ^4 \left(8
   \mathcal{L}^3+30 \mathcal{L}^2+57 \mathcal{L}+42\right)+\cdots.
   \ee
In Figure (\ref{fig:largesmallL}) we show the large and small $\mathcal{L}$
expansions for the $b_{0,1}(\mathcal{L})$ functions. We remark that for intermediate
values of $\mathcal{L}$ the two expansions nicely overlap.
For the next function $b_{p\ge 2}(\mathcal{L})$ we could not find a closed expression and $\mc L$
dependence away from $\mc L=0$ would seem unavailable.
Nevertheless, a numerical analysis  shows that, in terms of $q(\mc L)$, the functions
\be
\widetilde b_{p}(q) = \frac{\mathbb E(q)^{2p+1}}{1-q}\,b_{p}(q),
\ee
appears to be rather smooth for all $0\le q\le 1$. This remark suggests to re-define the following functions
\be
b_{p}^{[N,M]}(q) = \frac{1-q}{\mathbb E(q)^{2p+1}}\,{\rm Pade}[N, M, \widetilde b_{p}(q)],
\ee
where ${\rm Pade}[N, M, \widetilde b_{p}(q)]$ is the $[N,M]$ Pad\'e rational approximation of $\widetilde b_{p}(q)$
around $q=0$. In $b_{p}^{[N,M]}(q)$, $q$ is finally substituted by the exact solution of the second equation 
in (\ref{eq:b0-param}). Figure~(\ref{fig:pade}) shows the analysis of $b_{2}$ and $b_{3}$ which turns out to be quite 
robust with respect to the degrees of the Pad\'e approximants~\footnote{The same construction completely overlaps with the exact results in the case of $b_{1}$.}. Thus, this method provides a simple way to obtain the complete numerical profile of all the functions $b_{p}(\mc L)$.

\subsection{Turning on the sphere angle}

The case of a non zero $\theta$ angle is discussed in \cite{Gromov:2013qga}. Again the large coupling expansion of the \brem function can be arranged in the form:

\ba
\frac{1}{\theta \cot \theta} B_L &=& 
\frac{g}{\sqrt{\pi ^2-\theta ^2}}-\frac{6 L+3}{8 \left(\pi ^2-\theta ^2\right)}+
\frac{3 \left(\pi ^2 \left(6 L^2+6 L+1\right)-2 \theta
   ^2 L (L+1)\right)}{128 \pi ^2 g \left(\pi ^2-\theta ^2\right)^{3/2}}+\nonumber\\
&&
\frac{f_1}{512 \pi ^4 g^2 \left(\pi ^2-\theta
   ^2\right)^2}-\frac{f_2}{32768 g^3 \left(\pi ^6 \left(\pi ^2-\theta ^2\right)^{5/2}\right)}+O \left(\frac{1}{g}\right)^4
\ea
where $f_{1,2}$ are given by:
\ba
f_1&=& -3 \theta ^4 L \left(2 L^2+3 L+1\right)+6 \pi ^2 \theta ^2 L \left(2 L^2+3 L+1\right)+\pi ^4 \left(10 L^3+15 L^2+11 L+3\right)\nonumber\\
f_2&=&18 \theta ^6 L \left(5 L^3+10 L^2+7 L+2\right)-18 \pi ^2 \theta ^4 L \left(5 L^3+10 L^2+11 L+6\right)-
\nonumber\\ &&  6 \pi ^4 \theta ^2 L \left(55
   L^3+110 L^2+47 L-8\right)+9 \pi ^6 \left(10 L^4+20 L^3-22 L^2-32 L-7\right).
\ea
It is straightforward to check that even in this more general case all the involved polynomials satisfy (\ref{eq:Z2p})
order by order in the small $\theta$ expansion. We can therefore apply the same argument as in the $\theta=0$ case, and derive constraints like (\ref{eq:relation}).

\section{Conclusions}

The main results of this short paper is the $\mathbb Z_{2}$ symmetry (\ref{eq:reciprocity}) and its consequence 
(\ref{eq:relation}) relating the functions $b_{2n+1}(\mc L)$ to the even ones. 
Qualitatively, this is similar to what is found in the study of reciprocity invariance of various twist operators in $\mc N=4$ SYM. There, the anomalous dimension is a function of the operator spin $\gamma = \gamma(S)$. Under a non-linear change of variable, one  introduces a related function whose large $S$ expansion involves only integer
inverse powers of $S(S+1)$ at any order in the coupling constant. This means that roughly half of the terms in the 
 $1/S$ expansion can be expressed by the other half.
The relations (\ref{eq:relation})
are claimed to be valid in the gauge theory and it would be very interesting
to prove them by matrix model techniques as discussed  in \cite{Gromov:2012eu} in the case of  the classical term.
In principle, the discrete symmetry could also be investigated in the ${\bf P}\mu$-system \cite{Gromov:2013pga}
with the hope of being able to generalize it. 
Of course, the important open question is whether the functions $b_{p}(\mc L)$ do indeed reproduce
the various higher-loop semiclassical string computations. The formal equivalence of the expansion 
(\ref{eq:semiclassical}) with a semiclassical loop expansion in the string theory is intriguing, but 
only some terms could match the correspondence, beyond $b_{1}(0)$ that indeed passes the check. The simplicity of (\ref{eq:relation}) holding in the gauge theory
is remarkable and somewhat unexpected. An extension of the calculation in \cite{Drukker:2011za} could reveal 
whether it remains true in the dual string theory too.

\section*{Acknowledgments}

We thank Arkady Tseytlin, N. Gromov and Domenico Seminara for useful comments on the manuscript.

% \appendix

\bibliography{AC-Biblio}{}
\bibliographystyle{JHEP}

\end{document}